\documentclass[aps,pra,twocolumn,floatfix,showpacs,superscriptaddress]{revtex4}
\usepackage{amsmath,amssymb}
\usepackage{graphicx}
\newcommand{\W}{{\scriptscriptstyle W}}
\newcommand{\p}{{\scriptscriptstyle P}}
\newcommand{\V}{{\scriptscriptstyle V}}
\newcommand{\K}{{\scriptscriptstyle K}}
\begin{document}
\title{Finite-time quantum-to-classical transition for a Schr\"{o}dinger-cat state}   
\author{Janika Paavola}
\email[]{janika.paavola@utu.fi}
\homepage[]{www.openq.fi}
\affiliation{Turku Centre for Quantum Physics, 
Department of Physics and Astronomy, University of Turku,
FI-20014 Turun yliopisto, Finland}
\author{Michael J. W. Hall}
\affiliation{Theoretical Physics, Research School of Physics and Engineering, 
Australian National
University, Canberra ACT 0200, Australia}
\author{Matteo G. A. Paris}
\affiliation{Dipartimento di Fisica dell'Universit`a 
degli Studi di Milano, I-20133 Milano, Italia}
\affiliation{CNISM, Udr Milano, I-20133 Milano, Italy.}
\email[]{matteo.paris@fisica.unimi.it}
\author{Sabrina Maniscalco}
\email[]{s.maniscalco@hw.ac.uk}
\homepage[]{www.openq.fi}
\affiliation{SUPA, EPS/Physics, Heriot-Watt University, Edinburgh, EH14 4AS, UK }
\affiliation{Turku Centre for Quantum Physics, Department of Physics and 
Astronomy, University of Turku,
FI-20014 Turun yliopisto, Finland}
\date{\today}    
\begin{abstract}
The transition from quantum to classical, in the case of a quantum
harmonic oscillator, is typically identified with the transition from a
quantum superposition of macroscopically distinguishable states, such as
the Schr\"{o}dinger-cat state, into the corresponding statistical
mixture. This transition is commonly characterized by the asymptotic
loss of the interference term in the Wigner representation of the cat
state. In this paper we show that the quantum-to-classical transition
has different dynamical features depending on the measure for
nonclassicality used.  Measures based on an operatorial definition have
well defined physical meaning and allow a deeper understanding of the
quantum-to-classical transition.  Our analysis shows that, for most
nonclassicality measures, the Schr\"odinger-cat state becomes classical after a finite
time. Moreover, our results challenge the prevailing idea that more
macroscopic states are more susceptible to decoherence in the sense that
the transition from quantum to classical occurs faster.  
Since nonclassicality is a prerequisite for entanglement generation
our results also bridge the gap between decoherence, which is lost
only asymptotically, and entanglement, which may show a sudden death. 
In fact, whereas the loss of coherences still remains asymptotic, we emphasize that the transition from quantum to classical can indeed occur at 
a finite time. 
\end{abstract}
\maketitle
\section{Introduction}
Ever since the early days of quantum theory the gap between our
classical everyday reality and the quantum mechanical laws that govern
the microscopic world has been acknowledged. The Schr\"{o}odinger cat
gedanken experiment, in which a real cat is cleverly put in a
superposition of being alive and dead at the same time, illustrates the
seemingly paradoxical conclusions arising from the application of
quantum principles to macroscopic objects \cite{Schrodinger35}. 
\par
The prevailing explanation of the emergence of the classical realm from
the quantum is environment induced decoherence (EID) \cite{Zurek91,Paz93d}.
According to the EID description the reason why macroscopic quantum superpositions are
not observed in the classical world is the presence of the environment, which couples to all systems and effectively monitors quantum superpositions, inducing a collapse to the corresponding statistical
mixture of classical-like states (pointer states). Experiments observing
the quantum-to-classical transition, for Schr\"odinger cat-like states
of both light and massive particles, have been performed, e.g., in the
context of cavity QED and trapped ions, respectively
\cite{Wineland96b,Haroche96,Haroche08}.
\par
Quantum superpositions of macroscopically (or mesoscopically)
distinguishable states are sometimes called Schr\"odinger-cat states, in
the spirit of the original Schr\"odinger's thought experiment.
Typically, decoherence of such a quantum superposition state that leads to a statistical mixture, is identified with the transition from
quantum to classical, i.e., with the loss of the quantum features
initially possessed by the cat state \cite{Zurek91}.  According to one
of the earliest definitions, a state is classical if it can be expressed as
a statistical mixture of coherent states, i.e., if the $P$ function
\cite{Glauber63,Sudarhsan63,Glauber65,Glauber69} of the state is a
positive, well-defined probability distribution \cite{Glauber65}.
Although examples of nonclassical states in line with this original
definition exist \cite{Vogel08}, the $P$ function can also be highly
singular making its reconstruction very demanding.  Different
definitions and criteria for nonclassical states have been proposed in
the literature
\cite{Lee91,Barnett95,Klyshko96,Vogel00,Vogel02,Lewenstein05,Casamiglia05,Egger97}, 
also for multimode fields \cite{ths09,mir10,sol10,bri11}, and the decoherence process
has been analyzed extensively \cite{S1,S2}.
The different approaches are not equivalent, so the complete
characterization of nonclassical states, in particular a measurable
criterion that is both necessary and sufficient, does not exist, except for pure states \cite{Vogel02}. Besides their fundamental importance,
nonclassicality criteria are of key relevance also for quantum
technologies. Creating and revealing nonclassical states, e.g., is often
a prerequisite to generate entanglement for quantum information purposes
in all-optical setups \cite{Knight02,Wang02}.
\par
In this paper we consider five different definitions of nonclassicality
for a single mode of the quantum harmonic oscillator, paying special
attention to their physical meaning. We use these definitions to study
the quantum-to-classical transition, i.e., the dynamics of a
Schr\"odinger-cat state in presence of a dissipative environment inducing
decoherence. The nonclassicality indicators we deal with are: the peak
of the interference fringes of the Wigner function, the negativity of
the Wigner function, Vogel's noncalssicality criterion \cite{Vogel00},
the nonclassical depth \cite{Lee91,Barnett95}, and the Klyshko
criterion \cite{Klyshko96,DAriano99}. 
\par
As opposed to the definition based on the fringe visibility of the
Wigner function, which is the most widely used when describing the
quantum-to-classical transition, the other four criteria offer some
advantages. These definitions, indeed, have an operatorial
interpretation, connecting the transition process to a measurable
physical property. The interference fringes, on the contrary, 
are constructed mathematically assuming the full knowledge of the 
quantum state. In practice, however, any technique for quantum 
state estimation, including tomographic approaches, leads to a 
reconstruction of the density matrix within some confidence interval, 
so that the amount of nonclassicality is crucially influenced in a nonlinear 
way by the precision of the reconstruction technique \cite{LNP}.
\par
We find that, according to all the operatorial definitions, the
quantum-to-classical transition occurs at a finite time rather than
following an exponential decay, in accordance with the results found in Refs. \cite{JPADiosi,ActaDiosi,Marian} for the nonclassical depth and the negativity of the Wigner function. It is worth noting that, contrarily to entanglement, which is defined independently of the entanglement measure used, the concept of nonclassicality or quantumness of the state does depend on the nonclassicality criterion used. More precisely, even if for mixed states different entanglement measures may give different numerical values, they all agree on the minimum zero value indicating disentanglement. Therefore, entanglement sudden death does not depend on the specific measure of entanglement chosen.
On the contrary, as we will see, the loss of nonclassicality does depend on the criterion used to define nonclassicality. However, all operatorial definitions of nonclassicality show a similar behavior reinforcing the idea that the initial cat state loses its quantum character after a finite time, which we can identify with the maximum over the times the cat state becomes classical according to the different criteria.
\par
We also study how the
transition depends on the separation between the two coherent states of
the initial superposition, finding that the dependence of the
decoherence time from the separation, and therefore from the size of the
cat state, is not at all trivial and can be counterintuitive. It is
widely believed, indeed, that the more macroscopic the initial cat state
is, the faster is the quantum-to-classical transition. This is indeed
true for the fringe visibility criterion but, as we will see, using other 
criteria the situation changes drastically.
\par
The paper is organized in the following way. In Section II we introduce
the initial Schr\"odinger-cat state and the dynamics driving the
transition. In Sec. III we introduce the five different nonclassicality
conditions, we derive their environment-induced dynamics and, hence, we
single out and compare the characteristic features of the quantum-to-classical transition, according to each definition. Finally, in Sec. IV
we present concluding remarks and sum up the results.
\section{The system}\label{sec:sys}
Let us consider a quantum harmonic oscillator initially prepared in a
superposition of coherent states with opposite phases, i.e., in the
so-called Schr\"odinger-cat state,
\begin{equation}\label{cat}
|\Psi_{cat}\rangle=\frac{|\alpha\rangle+|-\alpha\rangle}{\sqrt{\cal N}}
\end{equation}
where $|\alpha\rangle$ denotes a coherent state and
$$\mathcal{N}=2[1+\mathrm{exp}(-2|\alpha|^2)]\,,$$ is
the normalization constant. For the sake of simplicity we will 
assume amplitude $\alpha$ real throughout the paper.  
We then assume that the oscillator interacts with a bosonic 
bath of oscillators at thermal equilibrium at temperature $T$.
In the Born-Markov approximation, and in the interaction picture,
the evolution of the system is governed by
the master equation 
\begin{align}\label{secME}
\frac{d\rho(t)}{dt}=&\,\gamma(n+1)\big[2a\rho(t)a^\dagger-a^\dagger
a\rho(t)-\rho(t)a^\dagger a\big]\\\nonumber &+\gamma n\big[2a^\dagger\rho(t)a-a
a^\dagger \rho(t)-\rho(t)a a^\dagger\big]\,, 
\end{align}
where $\rho$ is the density matrix of the quantum harmonic oscillator,
$\gamma$ the damping rate, $a$ and $a^\dagger$ the annihilation and
creation operators and $n$ the mean occupation number of the thermal
bath.
\par
All the nonclassicality criteria that we will use in the paper are based
on the quasiprobability distributions associated to quantum states.
These are the quantum analogs of the classical distribution functions
so, broadly speaking, any deviation from a classical probability
distribution is considered as a sign of nonclassicality.
The normalized quasiprobability distributions associated to the density
matrix $\rho$ are defined as the Fourier transforms of the
$s$-parametrized characteristic functions $\chi (\xi,s)$
\cite{Glauber69, Barnett95} 
\begin{equation}\label{generalizedW1}
W(\alpha,s)=\int\frac{d^2\xi}{\pi}\, \mathrm{e}^{\alpha\xi^*-\alpha^*\xi}
\,\chi(\xi,s), 
\end{equation}
where 
\begin{equation}\label{generalizedX}
\chi(\xi,s)=\mathrm{Tr}[\rho\,
\mathrm{e}^{\xi\hat{a}^{\dagger}-\xi^*\hat{a}}]\, \mathrm{e}^{\frac12 s |\xi|^2}\,.  
\end{equation}
The familiar $P$ function, Wigner function and Husimi $Q$ function are
obtained by choosing $s=1$, $0$ and $-1$, respectively.  These
distribution functions correspond to normal, symmetric, and antinormal
ordering of the creation and annihilation operators, respectively, and
they can be easily obtained from one another via convolution, i.e., for
$\tilde s < s'$, one has
\begin{align}\label{generalizedW}
W(\alpha,\tilde s)&= W(\alpha,s') \star G(s' -\tilde s, \alpha)  \notag
\\ &= \int d^2 \beta\,W(\beta,s')\, G(s'- \tilde s, \alpha-\beta)\,,
\end{align}
where
\begin{align}\label{gconv}
G(\kappa, \alpha) = \frac{2}{\pi\kappa}\, \mathrm{exp}\left(-2\,
\frac{|\alpha|^2}{\kappa}\right) 
\end{align}
Different distributions can be found useful for different tasks. The
Wigner function is often used to characterize nonclassicality because it
is bounded from above allowing experimental measurements. It is well
known, however, that defining nonclassicality in terms of the properties
of different quasiprobability distributions, e.g., the $P$ function or
Wigner function, does not yield equivalent results. 
The quantum-to-classical transition has been studied previously by
monitoring the time evolution of the interference peak in the Wigner
function representation \cite{Zurek91}. We include this approach in our
study, and compare it to four other possible ways to characterize the
quantum-classical border. Each approach has a different physical
interpretation, with different strengths and weaknesses. 
In the next section, we analyze such differences in an effort to obtain
insight into the emergence of classicality. To this aim we consider the
dynamics of different nonclassicality measures and study the time at
which the initial nonclassical state evolves into a classical one and
the dependence of such time from the relevant system parameters. 
\section{Loss of nonclassicality of the Schr\"odinger-cat state}
Monitoring the dynamics of the cat state as it evolves into a
statistical mixture is a natural way of studying the
quantum-to-classical transition since the initial superposition is not
an element of the macroscopic, classical reality, whereas the final
statistical mixture of minimum uncertainty coherent states is, the
latter states being the closest equivalent of a classical point in phase
space. The precise way of characterizing the transition leads to
different dynamical features and interpretations.
In the following we study analytically the time evolution of the peak of
the interference fringes of the Wigner function, the nonclassicality
depth, the negativity of the Wigner function, Vogel nonclassicality
criterion and the Klyshko criterion.
\subsection{Peak of the interference fringe}
A common way of monitoring the quantum to classical transition by using
the Wigner function is based on the time evolution of the highest point
of the interference term, characterizing the Schr\"odinger-cat state of Eq.
\eqref{cat}. Such a term is an indicator of the quantumness of the
superposition and hence its disappearance signals the transition to a
classical mixture. The presence of the interference peak can be
quantified via the fringe visibility function
\begin{align}
F(\alpha,t)&\equiv\mathrm{exp}(-A_{int})\nonumber\\
&=\frac{1}{2}\frac{W_I(\beta,t)|_{\mathrm{peak}}}{[W^{(+\alpha)}
(\beta,t)|_{\mathrm{peak}}W^{(-\alpha)}
(\beta,t)|_{\mathrm{peak}}]^{1/2}}\label{fringevis},
\end{align}
where $W_I(\beta,t)|_{\mathrm{peak}}$ is the value of the Wigner
function at $\beta=(0,0)$ and
$W^{(\pm\alpha)}(\beta,t)|_{\mathrm{peak}}$ are the values of the Wigner
function at $\beta=(\pm\alpha,0)$, respectively.  This is a widely used
signature for the emergence of classicality \cite{Paz93d} and it has
been experimentally monitored as well
\cite{Haroche96,Wineland96b,Haroche08}.
The time evolution of the fringe visibility for an oscillator initially 
prepared in a cat state and then evolving in a noisy channel reads as
follows
\begin{equation}\label{eq:fringe}
F(\alpha,\tau)=\mathrm{exp}\left[-2\alpha^2\left( 1-\frac{C_t^2}{1+ 2 D_t}
\right) \right], \end{equation}
where 
\begin{equation}\label{cd}
C_t =\mathrm{e}^{-\gamma t} \qquad D_t =n (1- \mathrm{e}^{-2\gamma
t})\,.
\end{equation}
The time evolution of the same quantity without the Markovian 
approximation (i.e., taking into account the memory effects of 
structured reservoirs) has been studied in \cite{Paavola10}.  
\par
As we can see from \eqref{eq:fringe}, the fringes disappear
asymptotically. If one then takes the peak of the interference fringe as
an indicator of nonclassicality, the quantum-to-classical transition
does not occur at a finite time.  No known operatorial expression,
however, can be given for the fringe visibility. Therefore it can be
seen as a calculational tool to describe decoherence, with no obvious
observable associated to it. Moreover, characterizing decoherence in
this way requires the full knowledge of the state density matrix,
obtained with complete tomographic measurements.  
\par
The dependence of the decoherence rate on the initial separation can be
seen from Eq. \eqref{eq:fringe}. This quantity is proportional to
$\alpha^2$, resulting in faster decoherence for more macroscopic initial
cat states. The explanation of the  emergence of the classical world
from the quantum one, according to environment induced decoherence, is
heavily based on this observation. More macroscopic states loose their
quantumness faster, and that is why we do not see any of the bizarre
effects predicted by quantum theory in our daily "macroscopic" life.
However, as we will see in the following subsections, this conclusion is
strongly dependent on the nonclassicality critierion considered, and
therefore cannot be used to corroborate the main traits of the quantum-to-classical transition, such as the dependence of the decoherence time
on the size of the system.
\subsection{Nonclassical depth}
Let us now focus on the nonclassical depth that can be obtained from the
properties of the generalized distribution functions introduced in Sec.
\ref{sec:sys}. The nonclassical depth was first introduced by Lee
\cite{Lee91} and, in a slightly different form, by L\"utkenhaus and
Barnett \cite{Barnett95} some years later. 
\par
The starting point is the $s$-parametrized quasidistribution function
given by Eq. \eqref{generalizedW}, with $s$ a continuous parameter.
Setting $s'=1$ in Eq. \eqref{generalizedW} one obtains an expression
giving  $W(\alpha,s)$ as a convolution of the $P$ function, 
\begin{equation}\label{LeeP}
W(\alpha,s)= P(\alpha) \star G(1-s,\alpha)
\end{equation}
Note that, for $s=1,0,-1$, $W(\alpha,s)$ coincides with the $P$, $W$ an
$Q$ functions, respectively. While the $P$ and the $W$ functions cannot
be generally considered proper distribution function, the $Q$ function
can, being always positive and regular. However, we note in passing that, even if the $Q$ function is always positive, its marginals are only approximate (broadened) position
and momentum variables. Hence its use as an indicator of classicality should be considered with care, as discussed in some detail, e.g., in Ref.  \cite{Milburn86}.
\par
The nonclassical depth of a given state is $$\eta=\frac12(1-\bar{s})$$ 
where
$\bar{s}$  is the largest value of $s$ for which $W(\alpha, s)$ is 
positive. For pure states $0 \le \eta \le 1$, while mixed states can
have any value of $\eta < 1$.  It was shown in Ref. \cite{Barnett95}
that for all pure states other than coherent squeezed states the
nonclassical depth is $\eta=1$, squeezed states have $0 \le \eta \le
1/2$, while coherent state have $\eta =0$, in accordance with the fact
that they are the closest analogue to classical states for the quantum
harmonic oscillator.
\par
In order to study the dynamics of the nonclassical depth $\eta$, for
the initial state of Eq. \eqref{cat}, we notice that the time evolution
of the $P$ function,  in presence of a dissipative thermal environment
leading to the master equation \eqref{secME},  can be written in a form
similar to Eq. \eqref{LeeP}. As we will see in the following, this
allows us to single out an analytic expression for the instant of time
$\tau_{P}$ which is an upper bound for the loss of nonclassicality.
The solution to the master equation \eqref{secME} can be written in
terms of the normally ordered characteristic function
$\chi(\xi, s=1) \equiv \Phi(\xi)$. Upon denoting by $\Phi_0(\xi)$
the characteristic function  at $t=0$ 
(i.e., the one of the initial cat state) we have that 
the characteristic function at time $t$ is given by
\begin{equation}\label{chit}
\Phi_t(\xi)=\Phi_0(C_t \xi)\,\mathrm{exp}(-D_t |\xi|^2),
\end{equation}
where the coefficients $C_t$ and $D_t$ are given in Eq. (\ref{cd}).
From Eq. \eqref{generalizedW1}, with $s=1$, and Eq. \eqref{chit} one obtains
\begin{align}
P_t(C_t \alpha) = 
\frac{1}{C_t^2}\,
&\int \frac{d^2\xi}{\pi}\, \Phi_0(\xi)\, 
\mathrm{e}^{- \frac{D_t}{C_t^2}\,|\xi|^2+\alpha\xi^*
-\alpha^*\xi }\,.
\label{eq:pt1} 
\end{align}
Remembering that the Fourier transform of a product of two functions is
equal to the convolution of the two corresponding Fourier transforms, we
can recast Eq. \eqref{eq:pt1} in the form
\begin{align}\label{hallresult}
C_t^2 P_t(C_t\alpha)&=  P_0(\alpha)\star G(1-s_t,\alpha) \\ 
&\equiv W(\alpha, s_t)\,, \notag
\end{align}
with 
\begin{eqnarray}
s_t=1-2v_t \,, \qquad v_t=D_t/C_t^2. \label{eq:stvt}
\end{eqnarray}
Thus, the master equation \eqref{secME} essentially turns the $P$
function of the initial state into the quasiprobability distribution
function $W(\alpha,s)$ of the initial state.  Indeed, at $t=0$, $s_0=1$
and the r.h.s. reduces to the $P$ function. As time increases $v_t$
increases and, correspondingly, $s_t$ decreases. An upper limit to the
time at which the state becomes classical is therefore given by the time
$t_{\p}$ at which $s_{t_{P}} = -1$, since in this case the $P$ function
of our initial state has become the $Q$ function, and therefore it is
positive. Note that $t_{\p}$ is an upper limit to the disappearance of
nonclassicality. Since the evolved state is always a mixed state,
indeed, $\bar{s}$  can be greater than  $s_{t_{P}}$. 
It follows via Eq. \eqref{eq:stvt} that the time $\tau_\p=\gamma t_{\p}$
is given by 
\begin{equation}
\tau_\p=\frac{1}{2 }\mathrm{ln}\left(\frac{1}{ n}+1\right)= \frac{\hbar \omega}{2k_B T},
\label{taup}
\end{equation}
in agreement  with Marian {\it et al.} \cite{Marian}, with $\omega$ the frequency of the harmonic oscillator, $k_B$ the Boltzmann constant, and $T$ the reservoir temperature.
After this time the state is classical, therefore $\tau_{\p}$ quantifies
the lifetime of the nonclassical Schr\"odinger-cat state. Note that this time is always
finite for any $n \neq 0$, in this sense we can talk about finite-time transition from quantum to classical for the Schr\"odinger-cat state. For smaller and smaller values of the reservoir
temperature, the lifetime of the cat state increases. 
\par
It is worth stressing that $\tau_{\p}$ is an upper bound to the
nonclassical depth, and therefore to the quantum-to-classical transition
time, for any initial nonclassical state since it corresponds to the
time at which the $P$ function of any initial state has evolved into a
positive distribution function, i.e., the $Q$ function, and since at all
times $t > 0$ the evolved state is a mixed state.
\subsection{Negativity of the Wigner function}
The negativity of the Wigner function has been used widely as a
nonclassicality definition, mostly due to the fact that the 
Wigner function is never singular, as opposed to the $P$ function,
and therefore it is possible to reconstruct it in an approximate way through 
quantum homodyne tomography. 
Recently it was shown that measuring merely two conjugate variables, 
instead of performing full state tomography, is sufficient to observe 
the negativity of the Wigner
function in a certified, error-free way \cite{Mari10}. 
\par
In the previous section we have seen that the master equation describing
the system dynamics, given by Eq. \eqref{secME}, essentially transforms
the $P$ function of the initial state into the generalized
quasidistribution function $W(\alpha,s)$ of the initial state, according
to Eq. \eqref{eq:pt1}. This equation describes also the evolution of the
initial Wigner function since at a certain time $\bar{t}$,
$s_{\bar{t}}=0$ and the dissipative channel has transformed the initial
$P$ function of our state into the $W$ function. It follows
straightforwardly that, an upper limit to the disappearance of
negativity of the Wigner function is given by the time $ t_\W$ such that
$s_{t_\W}=0$, i.e., following Eq. \eqref{eq:stvt},
\begin{equation} 
\tau_\W= \gamma t_\W=\frac{1}{2}\mathrm{ln}\left(\frac{1}{2n}+1\right)\,.  
\label{tauw}
\end{equation}
Note that, for high $T$-reservoirs, i.e., for $n\gg1$, $\tau_\W \approx
1/4n$ and $ \tau_\p \approx 1/2n = 2 \tau_\W$, indicating that the
negativity of the Wigner function disappears faster than the
nonclassical depth.
\subsection{Vogel criterion}
\subsubsection{1st order nonclassicality criterion}
The Vogel nonclassicality criterion states that a state is nonclassical if
there exist values of $u$ and $v$ such that 
\begin{equation}
  |\Phi(\xi)|  >1 \label{eq:condphi}
\end{equation}
for the normally ordered characteristic function, where $\xi=u+iv$.
In terms of the symmetrically ordered characteristic function 
$\chi(\xi,0)$ the condition reads
\begin{equation}\label{eq:cond}
|\chi(\xi,0)|>\chi_0(\xi,0) \equiv \mathrm{exp}\left(-\frac12 |\xi|^2\right)
\end{equation}
where $\chi_0(\xi,0)$ is the characteristic function of the
ground state of the system oscillator \cite{Vogel00}. 
It is worth noticing that the symmetric characteristic function can be
directly measured via balanced homodyne detection. Formulating a
criterion for nonclassicality in terms of the inequality \eqref{eq:cond}
therefore means that complete state tomography is not anymore necessary
to characterize the nonclassical status of a state. A single measurement
satisfying inequality \eqref{eq:cond} is sufficient, and maintaining a
stable relation between the local oscillator and the optical state
becomes unnecessary \cite{Shapiro02}.  This makes checking for the
nonclassicality of a state much simpler compared to full state
tomography. Experiments demonstrating the usefulness of the
nonclassicality criterion in Eq. \eqref{eq:cond} have already been
performed \cite{Shapiro02}.
\begin{figure}[h!]
\includegraphics[width=0.9\columnwidth]{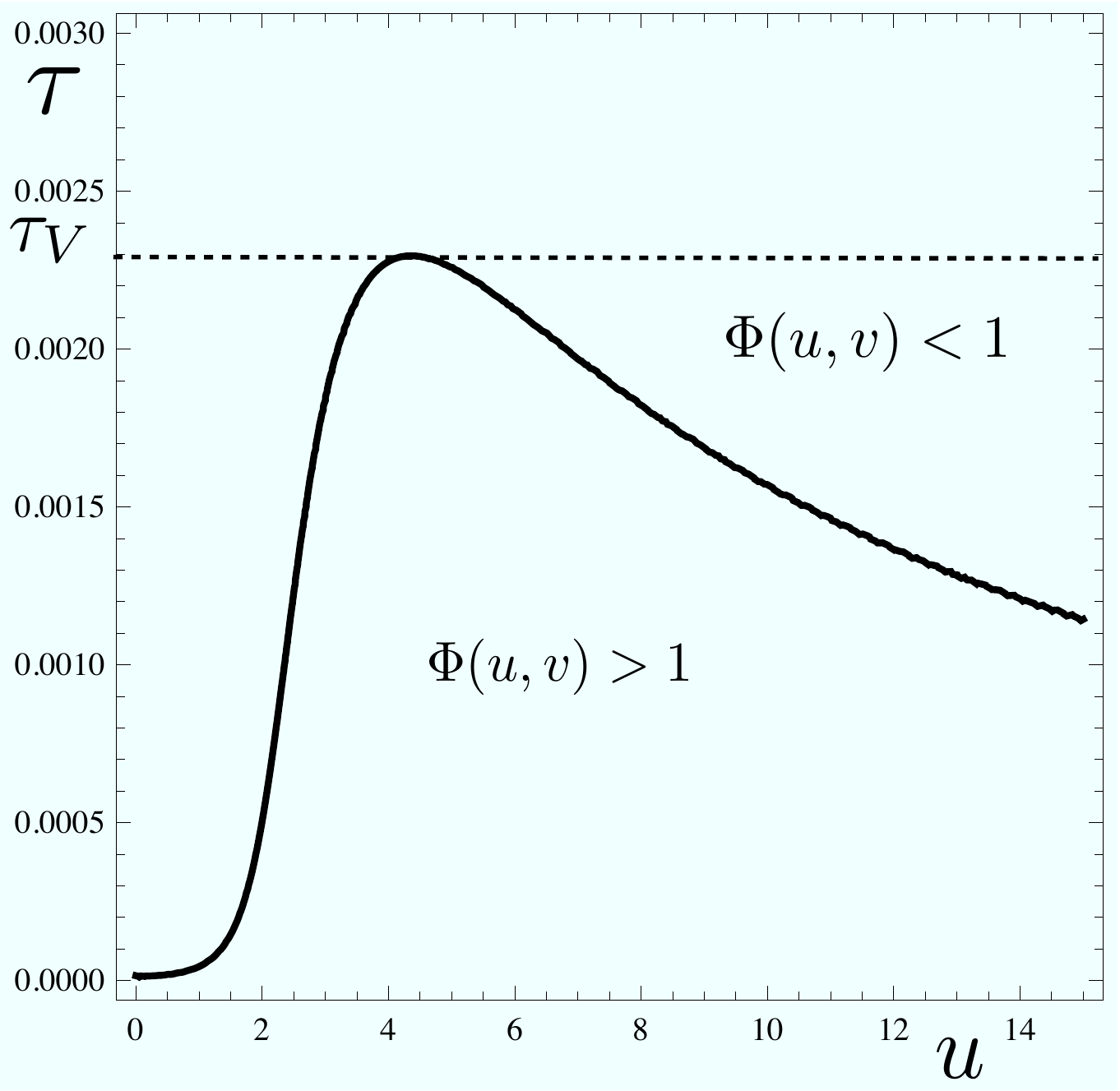}
\caption{Vogel nonclassicality condition as a function
of $\tau = \gamma t$ and  $u$, for $v=0 $, $n=100$ and $\alpha=2$. The
solid line corresponds to $ \Phi_t(u,0) =1$. The state is nonclassical
in the area under the curve. The time $\tau_\V$ is the time at which the
state becomes classical. } \label{fig:nonclassicality} \end{figure}
\par
However, some nonclassical states may not be captured by this
definition, as demonstrated by Di\'osi in Ref. \cite{Diosi00}. This
criterion is therefore sufficient but not necessary. The criterion was
later generalized by Richter and Vogel to give necessary and sufficient
conditions for nonclassicality \cite{Vogel02}. The new criterion
consists of an infinite set of conditions, considerably affecting its
practical usability. The original simple criterion of Eq.
\eqref{eq:cond} is still extremely useful as it can reliably and with
few measurements confirm an unknown quantum state as nonclassical.
For the initial cat state of Eq. \eqref{cat} the time evolution of the
normally ordered characteristic function
reads \cite{MQOB}
\begin{eqnarray}
 \Phi_t(u,v) &=& \frac{2}{\mathcal{N}}\, 
 e^{-D_t (u^2+v^2)}   \Big[ \cos(2C_t \alpha v)  \nonumber \\
&+& e^{-2 \alpha^2} \cosh(2C_t \alpha u) \Big]. \label{eq:phit}
\end{eqnarray}
Using Eq. \eqref{eq:phit}, we have investigated numerically the time
evolution of Eq. \eqref{eq:condphi} in the high $T$ limit finding that,
after a finite time, it is not satisfied anymore. Since $ \Phi_t(u,v)
\le  \Phi_t(u,0)$, we plot in Fig. \ref{fig:nonclassicality} the contour
line corresponding to $ \Phi_t(u,0) =1$. This contour line indicates the
transition from quantum to classical, according to Vogel first order
nonclassicality criterion. The dashed line indicates the time 
$\tau_\V (\alpha)$ after which the quantum property connected to the initially
macroscopically separated cat state, namely the one described by
condition \eqref{eq:cond}, is lost.
\subsubsection{Dependence on the size of the cat state}
\begin{figure}
\includegraphics[width=0.9\columnwidth]{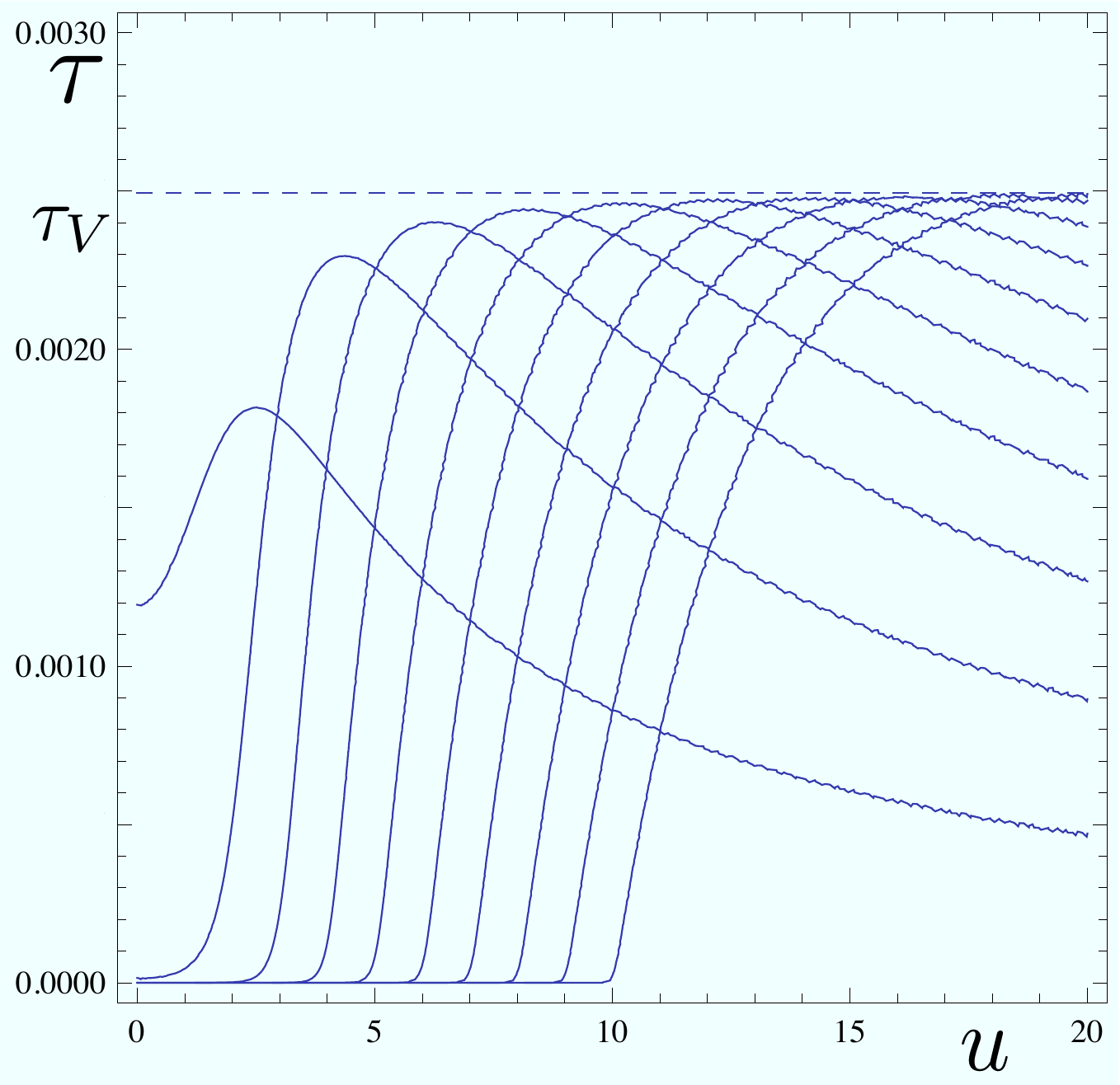}
\caption{Vogel nonclassicality condition $ \Phi_t(u,0)
=1$ as a function of $\tau = \gamma t$ and  $u$, for $v=0 $, $n=100$ and
initial separations $\alpha$ ranging from 1 to 10 (lines from left to
right, respectively). The dashed line marks the saturation time
$\tau_\V(\alpha \rightarrow \infty) $ after which the state is classical.
\label{fig:effectalpha}} \end{figure}

We now focus on the dependence of  $\tau_\V (\alpha)$ on the initial 
wave packet separation $\alpha$. In Fig. \ref{fig:effectalpha} we show how the
contour line  $ \Phi_t(u,0) =1$, indicating the loss of nonclassicality,
changes for increasing values of  $\alpha$. In more detail, we vary
$\alpha$ in unit step size from $1$ to $10$, corresponding to the curves
from left to right. 
Interestingly, the time $\tau_\V (\alpha)$ of loss of nonclassicality of the Schr\"odinger-cat state increases with the initial wave packet separation. This means
essentially that the more macroscopic the initial state is, the longer
it takes  to become classical.  This is in strong contrast with the
usual picture of emergence of the classical world from the quantum world
in terms of environment induced decoherence, according to which the more
macroscopic the cat state is, the faster is the quantum-to-classical
transition. Our results show that this is in fact only true for the peak
of interference fringes but not for other nonclassicality indicators, such as
the Vogel first order criterion.
\par
Another interesting feature shown in Fig. \ref{fig:effectalpha} is that
the time after which the nonclassicality condition \eqref{eq:cond}
ceases to be satisfied seems to saturate, possibly indicating an upper
bound for the onset of classicality for initially highly nonclassical
states. In fact it is possible to calculate such an upper bound
analytically noticing that
for $\alpha\rightarrow\infty$ one has
\begin{align} 
 \Phi_t(u,0) &\approx 
 e^{-D_tu^2} \notag \\ & 
 + \frac{1}{2} e^{-D_t(|u|-\alpha C_t/D_t)^2} e^{-\alpha^2(2-C_t^2/D_t)}. \label{eq:eqap}
\end{align}
From the equation above one can easily prove that a necessary and
sufficient condition for the state to be classical according  to the
first-order Vogel nonclassicality criterion, in the limit
$\alpha\rightarrow\infty$, is given by the equation 
\begin{equation}
C_t^2 \leq 2D_t. \label{eq:vogsat}
\end{equation}
Note that Eq.~\eqref{eq:vogsat} coincides with the equation defining
the time for the loss of nonclassicality, $\tau_\W$, in terms of the negativity of the
Wigner function.  Hence, 
\begin{equation}
\tau_\V (\alpha) \stackrel{\alpha\rightarrow\infty}{\longrightarrow} 
\frac{1}{2} \ln \left(1+\frac{1}{2n}\right)\equiv \tau_\W
\,. \end{equation}
\subsubsection{2nd order nonclassicality criterion}
The loss of nonclassicality, in the sense of $P$ function not being a
probability density, is not guaranteed by the condition \eqref{eq:condphi}.
Nonetheless, the use of Vogel first order nonclassicality criterion to
characterize the quantumness of a state has some benefits. In most
cases, indeed, it correctly identifies nonclassical states, the only
known exception being the example given by Di\'osi in \cite{Diosi00}.
Moreover, it is sufficiently simple to be of use in experiments and, in
the context of cat states and quantum-to-classical transition, it can be
used as a meaningful characterization of the dynamics since the initial
state satisfies Eq. \eqref{eq:condphi} but along the evolution the
inequality becomes invalid and hence, the state classical. In other
words, a property connected to the initially macroscopically separated
cat state, namely the one described in Eq. \eqref{eq:condphi}, has been
lost, and we choose to use this property as a characterization of the
quantum-to-classical transition.
\par
Although we argue that the nonclassicality criterion \eqref{eq:condphi}
could be used to indicate the finite-time quantum-to-classical transition, and do not aim to use in
this paper the infinite set of nonclassicality conditions by Richter and
Vogel \cite{Vogel02}, we have also numerically studied the time
evolution of the second order criterion which, in terms of the normally
ordered characteristic function, reads 
\begin{equation}
|{\Phi_1}|^2+|{\Phi_2}|^2+|{\Phi_{12}}|^2-2\mathrm{Re}{|\Phi_1 \Phi_2\Phi_{12}^*|}>1,
\end{equation}
where $\Phi_i=\Phi(u_1,v_1)$ and $\Phi_{ij}=\Phi(u_i+u_j,v_i+v_j)$.  We
have verified that the finite-time quantum-to-classical transition of the cat state occurs also in second
order but it takes a slightly longer time than in the case of the first
order condition.
\subsection{Klyshko  criterion}
The final nonclassicality criterion we are going to consider is based on
the work of Klyshko \cite{Klyshko96}. It takes the form of an inequality
involving terms from the photon number distribution of the mode under
investigation. Since photon number distributions may be effectively
reconstructed \cite{tomo,pcount} and in some cases also directly
measured \cite{burle,m}, this method has a clear experimental advantage.
Klyshko showed that an equivalence between a phase-averaged $P$ function,
\begin{equation}\label{phavep}
F(r)=\int_0^{2\pi}\frac{d\phi}{2\pi} P(r\mathrm{e}^{i\phi}),
\end{equation}
and an infinite set of inequalities concerning photon number
probabilities $p(n)=\langle n\vert \hat{\rho}\vert n\rangle$ exists,
providing a necessary and sufficient condition for nonclassicality in
terms of negativity of $F(r)$.  The simplest sufficient criterion for
nonclassicality  takes the form \cite{Klyshko96,DAriano99}
\begin{equation}
B(n)\equiv(n+2)p(n)p(n+2)-(n+1)[p(n+1)]^2<0.
\end{equation} 
For $F(r)$ to be negative, it is sufficient that this condition is
satisfied by just one non-negative integer number $n$.
The photon number probabilities can be obtained from
\begin{equation}\label{eq:B}
p(n,t)=\frac{1}{\pi}\int du\, dv\, \Phi_t(u,v)\,\chi_n(u,v),
\end{equation}
where $\Phi_t (u,v)$ is the characteristic function of the (evolved) 
cat state from Eq. \eqref{eq:phit} and $\chi_n(u,v)=\mathrm{exp}(-u^2-v^2)
L_n(u^2+v^2)$ is the anti-normally ordered characteristic 
function of Eq. \eqref{generalizedX}, with $s$=-1, for the Fock number 
state $|n\rangle$, $L_n(x)$ being the $n$-th Laguerre polynomials. For our 
initial cat state, the simplest condition showing the
nonclassicality is provided by negativity of $B(1)$. 
\begin{figure}[h!]
\includegraphics[width=0.48\columnwidth]{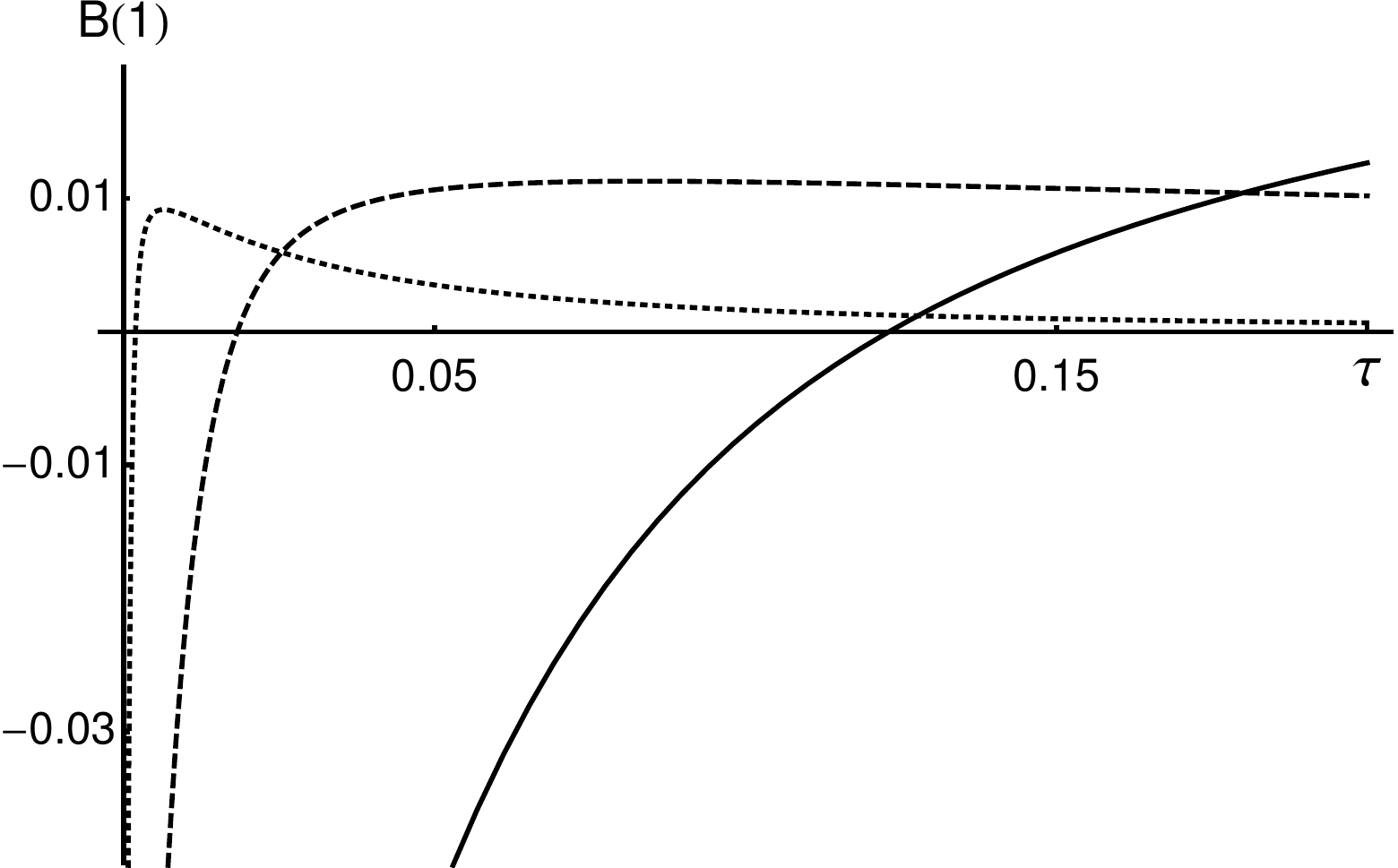}
\includegraphics[width=0.48\columnwidth]{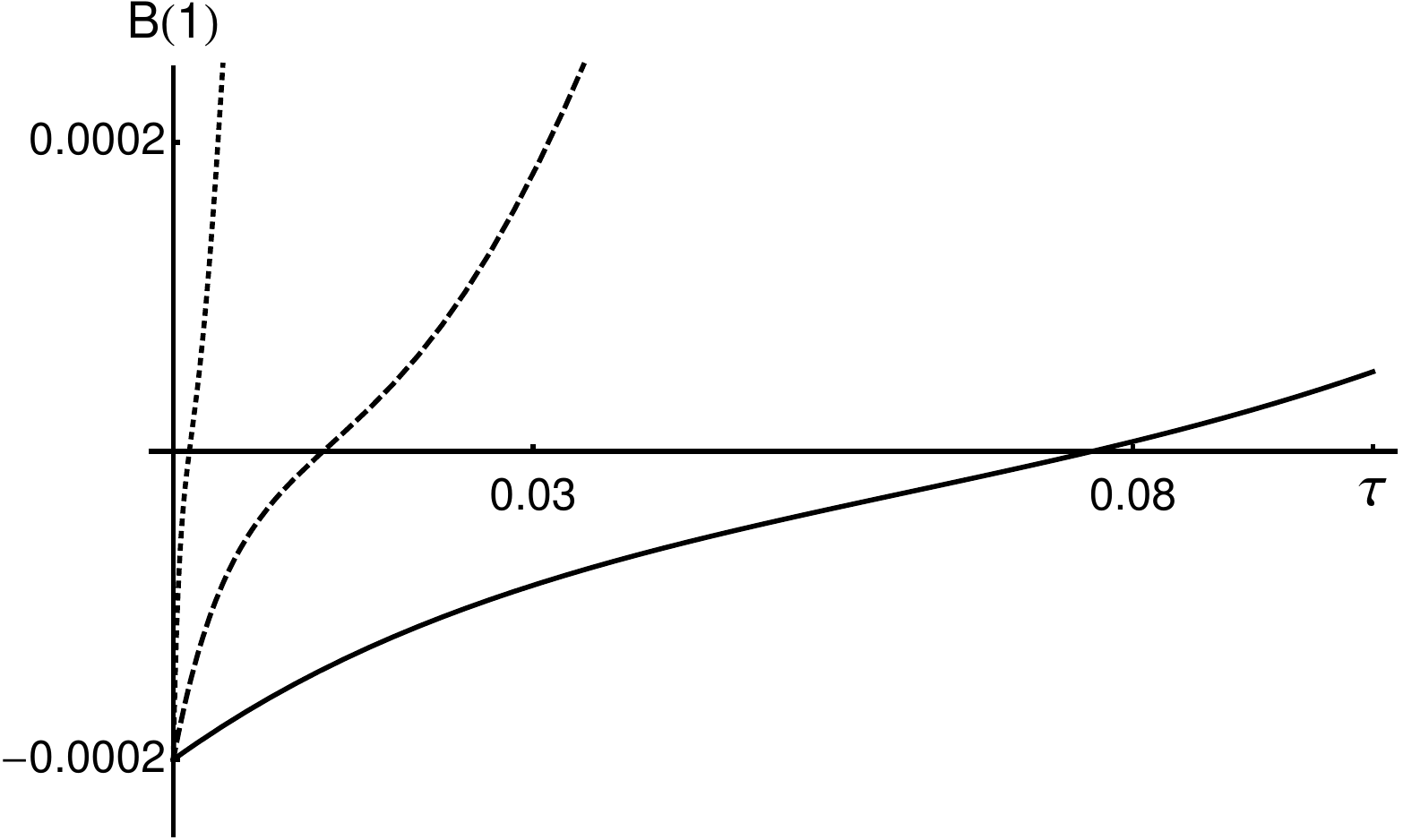}
\caption{The Klyshko quantity $B(1)$ as a function 
of the rescaled time $\tau=\gamma t$ for different
values of the thermal noise and different separation
amplitude (left: $\alpha=2$, right:$\alpha=3$). 
Nonclassical states turn into classical ones at a threshold 
time $\tau_\K$ depending on both the amplitude and the thermal noise.
In both plots solid lines are for $n=1$, dashed for $n=10$ and
dotted for $n=100$.}
\label{fig:B1}
\end{figure}
\par
\begin{figure}[h!]
\includegraphics[width=0.8\columnwidth]{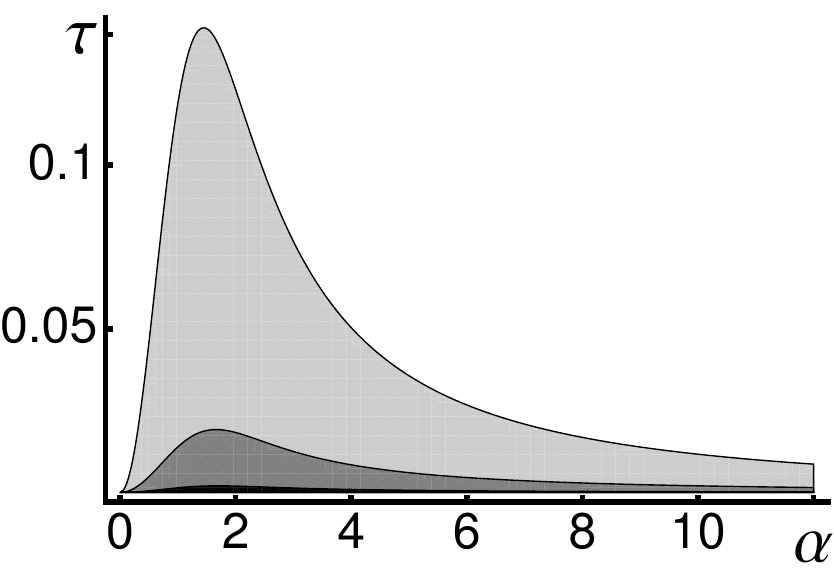}
\caption{The nonclassicality condition $B(1)<0$ is satisfied in 
the gray areas of the plot showing the transition time from 
quantum to classical as a function of the initial wave packet 
separation. Here $\tau=\gamma t$, and (from larger to smaller areas) 
$n=1, 10, 100$. The border of each gray area individuates the 
function $\tau_\K (\alpha)$.}\label{fig:B1alpha}
\end{figure}

In Fig.  \ref{fig:B1} we show the time evolution of $B(1)$ for different
values of the amplitude $\alpha$ and the thermal noise $n$. As it is
apparent from the plot, all the states exhibit a crossing from quantum
to classical state at a threshold time $\tau_\K$ which is a decreasing 
function of
the thermal noise.  The effect of initial separation, i.e., the function 
$\tau_\K (\alpha)$, is depicted in
Fig. \ref{fig:B1alpha} for different values of the thermal noise.
The nonclassicality condition $B(1)<0$ is satisfied in 
the gray areas of the plot, showing the transition time from 
quantum to classical as a function of the initial wave packet 
separation.
We see
that the classical domain is reached quite quickly for small and large
amplitudes, with weak dependence on the thermal noise, whereas an
optimum region of separation amplitudes exists ($\alpha\approx 2$) which
maximizes the survival of nonclassicality and introduces a strong dependence on the thermal noise.
\par
The behaviour of $B(n)$, for $n>1$, becomes increasingly difficult to obtain 
analytically. We have done some numerical comparisons and found indications
 that the nonclassicality thresholds obtained for higher order $B(n)$ 
are subsumed by that of $B(1)$. This was numerically confirmed for
$B(2)$ and $B(3)$.
\par
The dependence of the Klyshko criterion on the initial separation $\alpha$ is
qualitatively different from the ones predicted by all other criteria discussed in the paper.
As can be viewed from Fig. \ref{fig:B1alpha}, there exists a specific value,
$\alpha\approx 2$, that maximizes the time of nonclassicality for the
initial cat state. This is unique, since it implies that certain,
arbitrary, cat states are favored, in terms of the endurance of
quantumness. Actually, this is due to the structure of the quantity $B(1)$,
which is built from the overlap of the cat state characteristic function
with the characteristic functions of Fock states with small  values of $n$, all localized in the proximity of the phase-space origin. This circumstance, together with the fact that
higher order nonclassical tests seem to 
be subsumed by $B(1)<0$, suggest that the Klyshko criterion is not suitable 
to follow the time evolution of nonclassicality for highly separated 
superpositions.
\section{Discussion and conclusions}
In this paper we have addressed the problem of the quantum-to-classical
transition by examining the loss of crucial quantum properties in an
initial Schr\"odinger-cat state, which we identify with a quantum superposition of
two coherent states with opposite phases. Under the influence of a dissipative environment
the state decoheres and eventually reaches a state that can be
considered purely classical, namely,  a statistical mixture of the two
coherent states.
\par
We have shown that, depending on the measure of
nonclassicality considered, the
time at which one can say that the state has become classical, varies. In Table  \ref{tab:results} we list the different 
methods that we have compared in the paper, along with the numerical value of the time threshold 
$\tau$ after which the nonclassicality is lost for a given value of $\alpha$ 
and of the bath temperature $T$. We also summarize the dependence of the 
threshold value on the separation between the components of the quantum superposition and on $T$.
The only quantitatively different threshold time $\tau$, is related to the fringe visibility criterion, which gives an asymptotic
transition from quantum to classical. All the other measures predict that the quantumness of the state is lost after a finite time, i.e., there exists a sudden transition from quantum to classical for
the Schr\"odinger-cat state. It is notable that for the fringe
visibility no known operatorial interpretation exists, as far as the
authors are aware.
\par
By contrast, the quadrature characteristic function used in Vogel
nonclassicality criterion can be measured for freely propagating
radiation modes, cavity-field modes \cite{Meystre91} and the quantized
center-of-mass motion of a trapped ion in a harmonic potential
\cite{Wallentowitz95}. This last methods offers an operatorial approach
to the nonclassicality problem. It was shown in Ref. \cite{Wallentowitz95}
that the full state information of the vibrational motion of a trapped ion
can be obtained simply by monitoring the evolution of the ground state
occupation probability in a long-living electronic transition. 
We stress once more, however, that Vogel criterion, as given by Eq. \eqref{eq:cond}, does not
capture all nonclassical states, as Di\'osi demonstrated in
\cite{Diosi00}. However, since the criterion is satisfied for the
initial cat states, it singles out a property that belongs to such
superpositions. In this case the finite-time quantum-to-classical transition of the cat state coincides with the time at which the nonclassicality property associated to Vogel's first order criterion is lost.
\par
For the master equation \eqref{secME}, and for
finite bath temperatures, there exists always a transition from quantum
to classical, according to the original $P$ function criterion for
nonclassicality. This can be seen from the nonclassical depth.
Studying this quantity one sees that there exists always a specific time $t_P$ (or equivalently an amount
of noise that needs to be added to the system) after which the $P$
function of the initial state evolves into a classical $Q$ function. The explicit expression of $\tau_P$, given by Eq. \eqref{taup},
suggests the conjecture that classicality emerges
on a time scale inversely proportional to the effective temperature of the
environment, for very general systems and environments.
It is noteworthy in this definition that all pure initial
states, apart from coherent squeezed states, have the same threshold time for the quantum-to-classical transition 
(or, equivalently, they can withstand the same amount of noise added
before losing their quantumness). 
\par
The negativity of the Wigner function is another widely used indicator for
nonclassicality. However, it is well known that this quantity is
unable to identify all the states that are nonclassical according to the
$P$ function (squeezed states are a prime example). The popularity of the
Wigner function negativity  stems from the fact that the
Wigner function, unlike the $P$ function, can be measured with homodyne
detection. 
\par
Finally, the Klyshko criterion, which expresses the positivity of 
the phase averaged $P$ function in terms of the moments of the photon number
distribution, is the most sensitive of all the criteria
discussed here. It has the advantage of being
experimentally accessible since the photon number distribution, 
and in particular the probabilities needed to calculate
$B(1)$, may be reliably measured even by an on/off
detector \cite{pcount}. On the other hand, this quantity does not appear suitable
for superpositions of states with large separations. 
\par
The most important result of this paper is to challenge the current view
regarding the quantum-to-classical transition due to environment induced decoherence. Indeed, it
has been widely accepted that the more macroscopic the initial quantum superposition state is,
the faster is the decoherence and, hence, the transition from quantum to
classical. However, our results show that, for almost all the nonclassicality indicators, an increase in the
initial wave packet separation does not necessarily increase the time
after which decoherence has destroyed all nonclassical properties. 
The analysis of the  Vogel and Klyshko criteria, e.g., shows that the dependence on
$\alpha$ can be more complicated. In some cases, indeed, the transition time from quantum to classical can increase,
instead of decreasing, with the separation between the components of the superposition.
\par
What is conceptually interesting in our results is that they bridge the gap existing between decoherence and entanglement. Nonclassicality is a prerequisite for entanglement. However, the phenomenon of entanglement sudden death, discovered in 2001
\cite{Eberly02}, showed that entanglement can disappear completely after a  finite time while
decoherence, responsible for the loss of nonclassicality, 
decays only asymptotically \cite{Blatt}. The comparison of the dynamical features of several nonclassicality measures clearly shows that, while the loss of quantum coherences is indeed asymptotic, the quantum properties present in the initial state, which are defined by the measure chosen, disappear after a finite time.
%
\acknowledgments
We thank Erika Andersson and Andreas Buchleitner for useful discussions.
Financial support from the Emil Aaltonen foundation, the Finnish
Cultural foundation and the V\"ais\"al\"a foundation is gratefully
acknowledged.
\begin{widetext} $ $ 
\begin{table}[h!]
\caption{Threshold time for quantum-to-classical transition according to different indicators of nonclassicality for $\alpha=2$ 
and $n=100$. The last two columns summarize the dependence of the threshold value
on the cat state amplitude and on the mean occupation 
number of the thermal bath, i.e., on the temperature.}
\centering 
\begin{tabular}{clclclc} 
\hline 
Nonclassicality measure && threshold time $\tau$ && dependence on
$\alpha$ && dependence on $n$
\\ [0.5ex] 
\hline	
Klyshko criterion && 0.0019 && $\tau$ is maximum for $\alpha\approx 2$
&& decreasing with $n$, see Fig. \ref{fig:B1alpha}\\
Vogel criterion && 0.0023 && saturates with growing $\alpha$
&& decreasing with $n$ \\
Negativity of $W(\alpha)$ && 0.0025 && independent of $\alpha$
&& decreasing with $n$, see Eq. (\ref{tauw})\\
Negativity of $P(\alpha)$ && 0.0050 && independent of $\alpha$
&& decreasing with $n$, see Eq. (\ref{taup})\\
Fringe visibility && $\infty$ && proportional to $\alpha^2$
&&still asymptotic, converges faster, see Eq. (\ref{eq:fringe}) \\
\hline \end{tabular} \label{tab:results} 
\end{table}
\end{widetext}

\end{document}